\documentclass[11pt]{article}

\usepackage{epsfig}

\def\Journal#1#2#3#4{{#1} {\bf #2}, #3 (#4)}

\def\APPB{{\em Acta. Phys. Polon.} B}
\def\NPBPS{{\em Nucl. Phys.} B({\em Proc. Suppl.})}
\def\IJMPA{{\em Int. J. Mod. Phys.} A}
\def\JHEP{{\em JHEP}}
\def\NPB{{\em Nucl. Phys.} B}
\def\PLB{{\em Phys. Lett.}  B}
\def\PRL{\em Phys. Rev. Lett.}
\def\PRD{{\em Phys. Rev.} D}
\def\be{\begin{equation}}
\def\ee{\end{equation}}

\begin{document}

\hbox to\hsize{%
  \vbox{%
        }\hfil
  \vbox{%
        \hbox{MPI-PhT/37-2001}%
	\hbox{TPJU/7-2001}%
        \hbox{September, 2001}%
        }}

\vspace{1cm}
\begin{center}
\Large\bf
 LATTICE STUDY OF THE SIMPLIFIED MODEL OF M-THEORY FOR
LARGER GAUGE GROUPS

\vskip5mm
\large P. Bialas${}^a$  and J. Wosiek${}^b$\footnote{Invited talk presented at the
Sixth Workshop on Non-Perturbative QCD, American University of Paris,
Paris,  June,  2001.}

\vspace{3mm}
\small\sl
${}^a$ Inst. of Computer Science\\
Jagellonian University, Nawojki 11,\\
30-072 Krakow, Poland\\
${}^b$ M. Smoluchowski Institute of Physics,\\
 Jagellonian University, Reymonta 4, \\ 
30-059 Cracow, Poland\\ 
E-mail: wosiek@th.if.uj.edu.pl


\end{center}
\vspace{5mm}
\begingroup \addtolength{\leftskip}{1cm} \addtolength{\rightskip}{1cm}
\subsection*{Abstract}

Lattice discretization of the supersymmetric Yang-Mills quantum mechanics is discussed.
First results of the quenched Monte Carlo simulations, for D=4 and with higher gauge groups
( 3 $\le $ N $\le $ 8 ), are presented. We confirm an earlier (N=2) evidence that the system reveals 
different behaviours at low and high temperatures separated by a narrow transition region. 
These two regimes may correspond to a black hole and 
elementary excitations phases conjectured in the M-theory. Dependence of the "transition 
temperature" on N is consistent with 't Hooft scaling and  shows a smooth saturation of lattice 
results
towards the large N limit.  Is not yet resolved if the observed change between the
 two regimes corresponds to a genuine phase transition or to a gentle crossover .
A new, noncompact formulation of the lattice model is also proposed and its advantages 
are briefly
discussed.

\endgroup
\vspace{1cm}


\section{Yang-Mills quantum mechanics and its lattice formulation}
The hypothesis of  M-theory was a culmination of the recent developments
in nonperturbative string theories.  In the unified scheme, which emerged, all known five string 
theories are dual images of a single theory, which also contains 
eleven dimensional supergravity (for a recent review see e.g. \cite{SEN}).
Even though it is not yet known with the full precision,
 the M-theory has a remarkable
potential to unify all interactions and particles. In particular
it may offer a topological explanation of such fundamental
features
 as three families and fractional charges. It may lead to a standard model gauge group
with $\cal{N}$=1 supersymmetry. It provides understanding of the
Bekenstein-Hawking entropy puzzle and much more. Due to the proposition of Banks,
Fishler, Susskind and Shenker \cite{BFSS} ,  M-theory became amenable to yet more
 quantitative studies. According to BFSS the 
spectrum of M-theory is equivalent to that of a supersymmetric
Yang-Mills quantum mechanics (SYMQM). The latter is obtained from the
dimensional reduction of the 10 dimensional supersymmetric
Yang-Mills theory to the one (time) dimension. Such a quantum mechanical system
can be quantitatively studied by a host of  nonperturbative
methods. One of them, which  provides steadily increasing understanding of nonperturbative 
phenomena, is the lattice approach.  Therefore we have
constructed the Wilson discretization of the above quantum
mechanics and proposed to investigate it with the standard lattice
methods \cite{JW}. The ultimate goal is to study D=10 SYMQM for
the large size of the SU(N) matrices. However, even for this
relatively simple, one dimensional quantum mechanical system this
is still a rather complex task. One of the main difficulty is posed  by the complex
fermionic determinant (and pfaffian) for addjoint fermions in D=10.
Another, although less severe  one, is the  time consuming
Monte Carlo calculation with large nonabelian matrices. On
the other hand the system can be simulated at present in other
regions of the parameter space $D,N,N_f$, ($N_f$ being the number of fermions )
 and such a study may
provide a relevant information about its general structure. Therefore we
have decided to set up a systematic lattice survey of SYMQM
beginning with the simplest case of $D=4, N=2, N_f=0$ and
gradually extending it as far as possible towards the BFSS limit.
 In this talk I will  report on the second step along this program, namely the first results for 
 higher N will be presented.

     Supersymmetric Yang-Mills quantum mechanics \cite{CH,UPP,HS}
      and its zero dimensional counterpart \cite{IKKT,KOP} have been intensively studied.
      Although
the exact solution is still not available, many results are known and a) can be tested and extended, 
and b)
provide us a guidance from a "simple corner" of parameter space to the ultimate "BFSS corner".

       The action of the SYMQM reads
\begin{equation}
 S=\int dt \left({1\over 2} \mbox{\rm Tr} F_{\mu\nu}(t)^2
                     +\bar\Psi^a(t){\cal D}\Psi^a(t) \right). \label{QM}
\end{equation}
where $\mu,\nu=1\dots D$, and all fields are independent of the space coordinates $\vec{x}$.
 The  supersymmetric fermionic partners
 belong to the addjoint representation of SU(N). The discretized system is put
on a $D$ dimensional hypercubic lattice
$N_1\times\dots\times N_D$ which is reduced in all space directions to $N_i=1$,
$i=1\dots D-1$. 
Gauge and fermionic variables are assigned to links and sites of the new
elongated lattice in the standard manner.
This geometry has two straightforward consequences. First, all space derivatives
in bosonic and in fermionc terms vanish, $\partial_i \rightarrow 0 $. Second, there is a new 
class of gauge invariant observables, namely $Tr( A_i(t)^2)$ or $Tr( U_i(m))$,  since a local 
gauge transformation of the spatial components $A_i(t)$ reduces to a simple similarity 
transformation.
 The gauge part of the action has the usual form
\be
S_G=-\beta
\sum_{m=1}^{N_t} \sum_{\mu>\nu}
{1\over N} Re( \mbox{\rm Tr} \, U_{\mu\nu}(m) ),
\label{SG}
\ee
with
\be
\beta=2N/a^3 g^2,  \label{beta}
\ee
and $U_{\mu\nu}(m)= U_{\nu}^{\dagger}(m)U_{\mu}^{\dagger}(m+\nu)
     U_{\nu}(m+\mu)U_{\mu}(m) $,
$U_{\mu}(m)=\exp{(iagA_{\mu}(a m))}$, where $a$ denotes the
lattice constant and $g$ is the gauge coupling in one dimension. 
The integer time coordinate along the lattice is $m$. Periodic
boundary conditions $U_{\mu}(m+\nu)=U_{\mu}(m)$, $\nu=1\ldots
D-1$, guarantee that Wilson plaquettes $U_{\mu\nu}$ tend, in the
classical continuum limit, to the appropriate components
$F_{\mu\nu}$ with the space derivatives absent. 

Another simple property of the one dimensional gauge system
shows up in Eq.(\ref{beta}). The gauge coupling $g$ has a dimension,  and as a consequence
its numerical value provides directly a scale to all phenomena occuring in this system. 
In a language familiar from the four dimensional QCD: the dimensionless
lattice coupling $\beta=\beta(a) $ is running according to Eq.(\ref{beta}) , and $g$ plays
a role of $\Lambda_{QCD}$. 

In the above formulation
the projection on the gauge invariant states is naturally implemented.

Eq. (\ref{SG}) was the basis of the MC simulations 
for a simplified SU(2) model ({\em see} Sect.(2.1)).
Now some results for higher gauge groups $N<9$ are 
also available and will be discussed in Sect.(2.2).

\section{Results}    
\subsection{SU(2)}
One of the remarkable features of the M-theory
is the explanation of the problem of the Bekenstein-Hawking entropy\cite{STRO}
in terms of the elementary excitations. 
Moreover, the theory
predicts existence of at least two phases: a low
temperature, "black hole", phase and a high temperature phase
described by the elementary D0 branes \cite{MAR} \footnote{ The full
phase structure of the M-theory is  much more
complex, {\em op. cit.}}. Therefore a natural question emerges if the
simplified model (i.e. SYMQM), considered at finite temperature,
 possesses any nontrivial phase
structure \cite{KAB}.  It is well established that
QCD (or pure Yang-Mills theory)
 has two different phases (e.g. confinement and deconfinement).
  Since the action (\ref{SG}) is basically QCD-like, one
 might naively expect that the dimensionally reduced model
 may indeed exhibit similar phenomenon. 
 On the other hand  the one dimensional model with local interactions cannot
have a phase transition for finite N. However at infinite N the sharp singularity may occur \cite{GW}.
 It follows that, contrary to the first expectations, the mechanism which may 
 eventually lead to such a transition will have to be rather different that that for, 
 {\em e.g.}, $N=3$ QCD.  Hence the problem is even more interesting, since a) it is relevant
 for the M-theory builders and b) the nature of the transition (if any) is open.
 
 In Ref.\cite{JW} we have studied the phase structure of the quenched model
 with D=4, N=2  and $N_f=0$.  Since N plays a role of 
 the volume, we expected to see  some signatures of a phase change for finite
 and even small N \footnote{It is well known from statistical physics that the singularities
 (in the temperature) associated with the phase transition, develop gradually and smoothly 
 with increasing the volume and show up ( as a broad peaks for example) even for small systems
 \cite{SW}. }. 

 To check this we have measured the distribution
of the trace of the Polyakov line
\be
    P={1\over N}  \mbox{\rm Tr} \left( \prod_{m=1}^{N_t} U_D(m)
\right).  \label{poly}
\ee
which is a very sensitive determinant of the two phases in QCD. Similarly to lattice pure gauge
system,
symmetric concentration of the trace around zero indicates the low temperature phase
with $<P>=0$ (here a "black hole" phase) while clustering around $\pm 1$ (or, for arbitrary N,
 around the elements of $Z_N$) is characteristic of the high temperature phase
(here the elementary 0-brane phase). 
 It follows from Eq.(\ref{beta}) and $N_t=1/Ta $ that $\beta=2N N_t^3 T^3 $, hence the lattice
  coupling provides, 
 at fixed $N_t$ ,  a direct measure of the physical temperature $T$ \footnote{Up to the
 finite $a$ corrections.} .
 It was found that the
distribution of (\ref{poly}) is indeed changing from a convex to a
concave one at some finite value of $\beta=\beta_c(N_t)$ . Moreover
the dependence of $\beta_c$ on $N_t$ is
 consistent with the canonical scaling. The best fit to Monte Carlo results gave
$\beta_c(N_t)=(0.17\pm 0.05)N_t^{(3.02\pm 0.33)}$ in good
agreement with the expected dependence $\beta_c\sim N_t^3$. This
means that in the continuum limit the transition occurs at
{\em finite} temperature $T_c$, which can be written as $T_c=A_N (g^2 N)^{(1/3)}$
with $A_2=.28\pm 0.03$.   


We have also measured the average size of the system
$R^2=g^2\sum_a (A_i^a)^2$  for different
values of the temperature. It shows entirely different behaviours at low and high 
temperatures with a narrow transition region.
The
pseudocritical temperature determined in this way  is consistent
with the one obtained above from the study of the Polyakov line.
Our results also agree qualitatively with
the mean field calculation at large $N$ \cite{KAB}.
Obviously there is a long way
between $N=2$ and the BFSS limit $N\rightarrow\infty$ and it was essential
to extend this study to the larger groups. 

\subsection{SU(  3  $\le $ N  $\le $  8)}    
Recently we have done Monte Carlo simulations with the intermediate size groups,
still for $N_f=0$ and $D=4 $ \cite{BW}.  We used the standard Metropolis algorithm employing
all independent moves in the SU(N) manifold. For $N>2$ the trace P
in Eq.(\ref{poly}) is complex. The condition det(U)=1 constraints $P$ to lie inside the
"N-star" in the complex plane ({\em c.f.} Fig.1).  New simulations entirely confirm the structure
found in the SU(2) case. In the low temperature phase  the distribution 
is symmetric and peaked around $P=0$ which results in $<P>=0$. At high temperatures
P's are expected to concentrate around the elements of $Z_N$. 
A sample of our results for N=3,5,8 and at high and low temperatures 
$(\beta\sim T^3)$ is shown in Fig.1. Indeed one clearly sees the existence of both
 regimes. Moreover, in the high temperature region, the system has a tendency to be stuck
in one of the $Z_N$ minima. Correspondingly the thermalization time grows strongly with N.
This is a typical indication of the spontaneous symmetry breaking which might occur at
 $N\rightarrow \infty$ leading eventually to a non-zero value of the order
parameter~$<P>$.

\begin{figure}[t]
\epsfxsize=30pc
\epsfbox{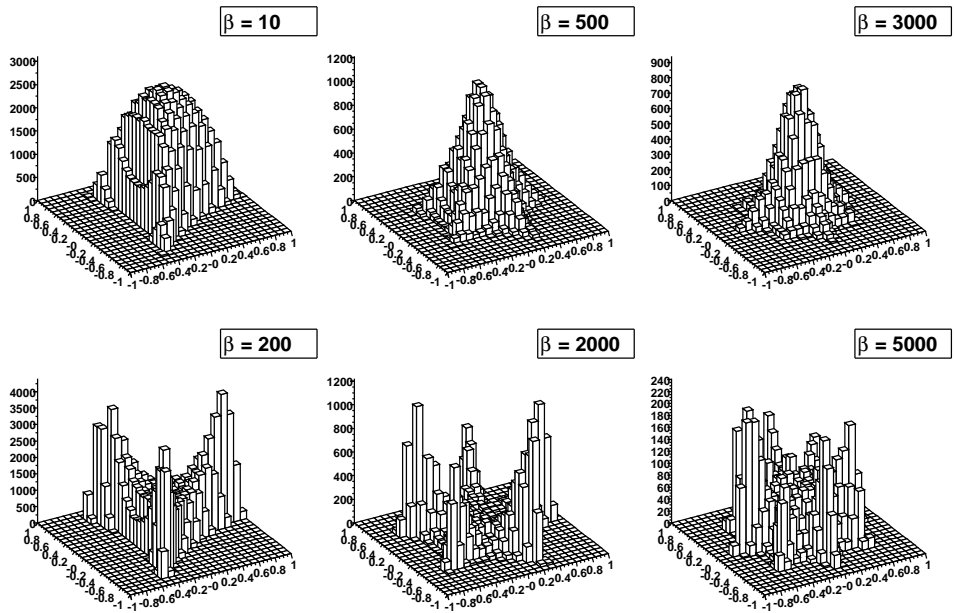}
\caption{Distribution of the Polyakov line in the low  and high 
temperature phases (upper and lower rows respectively) of the quenched Yang-Mills 
quantum mechanics 
with the SU(3), SU(5) and SU(8)
gauge groups.  \label{fig:radish}}
\end{figure}

We cannot determine at present the nature of the transition separating the two
phases. In fact it is also possible that this is not the phase transition at all, but just a gentle
crossover between the high and low temperature regimes.  More extensive simulations 
are required to answer this question. Nevertheless, our data show unambiguously that 
the system behaves differently in both regions. For example, the dependence of the size
of the system on the temperature is definitely different, and the change in the behaviour
occurs at the same T where the distributions in Fig. 1 change their shapes.
 Moreover, the characteristic temperature separating the two regions (loosely called phases) 
 is physical in the sense that it is finite in the continuum limit. This follows simply from 
 the $N_t^3$ scaling of the lattice coupling $\beta_c$. 
 
       With present data one can study, for the first time, the $N$ dependence of the transition
temperature. It follows from Eq.(\ref{beta}) and $N_t=1/Ta$, that the 't Hooft scaling 
$T_c\approx (g^2 N)^{1/3}$ implies
that the lattice coupling $\beta_c\approx N^2$ at fixed $N_t$. This prediction is tested 
in Fig.2, where
$\beta_c/N^2$ is plotted versus N. The values of $\beta_c$ were determined for each $N$,
 analogously to Sect.2.1, as the couplings for which the distributions, Fig.1, change their shape. 
The errors represent the subjective uncertainty of such an estimate. 
 Indeed a nice saturation of the N dependence is observed in Fig.2 confirming the 't Hooft 
 scaling of $T_c$. The solid line is a fit of the $1/N^2$ correction, and the flat solid line
  indicates the  fitted asymptotic value.  The data for the first two $N$ require yet higher order 
  corrections, while
  simulations for SU(8) give the asymptotic value with less than  15\% accuracy. 
  
  It is interesting to compare above results with the N dependence found in the 
  full, spatially extended Yang-Mills theory \cite{TEP} where the $1/N^2$ correction
  seems to work well even for N=2,3 in some cases. It would be instructive to repeat
  their study for the finite temperature phenomena.

\begin{figure}[t]
\epsfbox{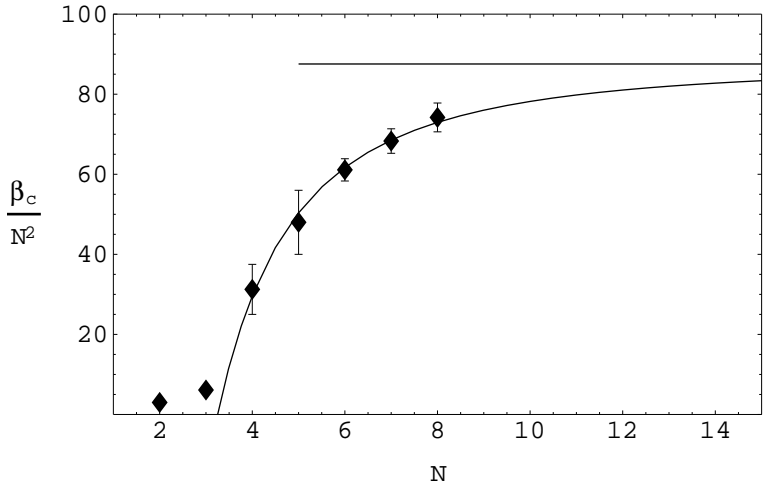}
\caption{Dependence of the transition lattice coupling on the number of colours $N$ at fixed 
time extent $N_t=4$.}
\end{figure}

\section{Noncompact formulation}
In this section we propose an alternative discretization of the 
continuum model (\ref{QM}), which turns out to be easier to simulate numerically.
The idea is to leave the $D-1$ spatial components of the gauge fields noncompact, while
keeping the $D$-th component compact. To this end split the continuous action 
of the reduced theory into the kinetic and potential terms. 
\begin{equation}
\begin{array}{rcl}
S&=& S_{kin}+S_{pot},  \\[4pt]
S_{pot}&=&{1\over 2 g^2 }\int dt Tr ({\bf X}^i {\bf X}^k)^2, \\
S_{kin} & = &{1\over  g^2 }\int dt Tr (D_t X^i) ^2. 
\\ & &
\end{array}\label{eq:spa}
\end{equation}
where the sums over $i$ and $k$ are
implied and $D_t$ denotes the covariant derivative along the time direction
 After discretizing the time we obtain
\begin{equation}
\begin{array}{rcl}
S_{pot}&=&{a\over 2 g^2 }\sum_m  Tr ({\bf X}^i {\bf X}^k)^2, \\
S_{kin} & = &{1\over  a g^2 }\sum_m Tr (\Delta {\bf X}^i) ^2,  
\\ & &
\end{array}\label{eq:disc}
\end{equation}
with the $D-1$ noncompact matrix coordinates ${\bf X}^i(m)$ defined at a discrete time intervals 
$t_m=a m$. The  covariant finite difference along the time direction
\begin{equation}
\Delta {\bf X}^i(m+1)={\bf X}^i(m+1)-U(m+1,m){\bf X}^i(m)U(m,m+1).
\end{equation}
 takes into account the parallel transport between adjacent lattice cites.
The system now has $D-1$ noncompact coordinates ${\bf X}^i$ and one compact 
degree of freedom corresponding to the timelike link
 $U(m+1,m)\equiv U_D(m)$.
The discretized action is invariant under  the same local gauge transformations
as the compact version (\ref{SG})
\begin{equation}
\begin{array}{rcl}
{\bf X}^i(m) & \rightarrow & V^{-1}(m){\bf X}^i(m)V(m), \\
U(m+1,m) & \rightarrow & V^{-1}(m+1) U(m+1,m)V(m),  
\\ & &
\end{array}\label{eq:spb}
\end{equation}
As a consequence of the dimensional reduction there is no inhomogeneous 
term in the gauge transformation of the spatial coordinates ${\bf X}^i$. Also,
they are rotated by {\em the same} element
of the gauge group $V(m)$ on both sides, since the space like links on $N_i=1$ lattice , 
Eq.(\ref{SG})  close back onto themselves.  

With the noncompact action (\ref{eq:disc}) one can readily perform simulations 
 closer to the continuum limit than those done with the fully compact variables.
Also higher N's can be easier achieved.
This is because the lattice coupling $\beta=2N/a^3g^2$ diverges strongly at small $a$ and this
causes very severe critical slowing down  in the compact case. The problem is even
more difficult for larger $N$. On the other hand the effective coupling in the noncompact 
formulation (\ref{eq:disc}) diverges only as $1/a$ which alleviates the above difficulty.

We have done some exploratory simulations with the new formulation for a range of
$N$'s $(  2  \le N \le 9 )$ and for different dimensions of unreduced theory $(  4 \le D \le 10)$.
All simulations in all these dimensions confirm existence of the two different regimes
of the model in full analogy with the compact, $N=2 $ case. We have encountered no difficulty
extending these study to higher D. In fact we have found a steady decrease of the 
average size of a system (defined per one link) with D, in both phases, in agreement 
with the mean field predictions \cite{KAB}.

\section{The future}
Lattice simulations provide a new approach to a quantitative study of the Yang-Mills 
quantum mechanics and possibly to the M-theory. Preliminary results are encouraging, but
a lot remains to be done.  Simulations of the quenched model work for all interesting
values of the dimension $D$ and are feasible for a range of N. Recent results give us a rough idea
how the large N limit is approached and where the asymptotics sets it. All quenched simulations 
performed up do date indicate existence of the two regions at finite temperature. This intriguing 
correspondence with the predictions of the M-theory should be (and can be) further quantified.
Of course, the next step is to include the dynamical fermions. This can be done by a brute force
for D=4 and for first few N's at D=10.  The one dimensional nature of the system
should help considerably.   For higher N, at D=10, we face the problem of the
complex pfaffian and new ideas are needed.   Finally, the full potential of the small volume
approach \cite{LVB,VB} may provide an important insight into the problem and should be  
applied to these  supersymmetric systems.

\section*{Acknowledgements}
JW thanks the organizers of this Workshop for the invitation and support. JW also thanks the 
Theory Group of the Max-Planck-Institute in Munich, where this work was completed, 
for their hospitality and support.
This research is supported by the Polish Committee for Scientific Research
under the grant 2 P03B 019 17.

\end{document}